\documentclass[aps,prl,twocolumn,preprintnumbers,amssymb,showpacs,superscriptaddress,floatfix]{revtex4-1}
\usepackage{amstext,amssymb}
\usepackage{amsmath}
\usepackage{amsfonts}
\usepackage{xcolor}
\usepackage{graphicx}% Include figure files
\usepackage{dcolumn}% Align table columns on decimal point
\usepackage{bm}% bold math
\usepackage{tikz}
\usetikzlibrary{decorations.pathmorphing,patterns}
\usepackage{braket}
\usepackage{color}
\usepackage{bbold}

\DeclareMathOperator{\Tr}{Tr}

\begin{document}
	\def\mean#1{\left< #1 \right>}
	%\preprint{}
	
	\title{Dissipative charging of a quantum battery}
	
	\author{Felipe Barra\footnote{fbarra@dfi.uchile.cl}}
	\affiliation{Departamento de F\'isica, Facultad de Ciencias F\'isicas y Matem\'aticas, Universidad de Chile, Santiago, Chile}
	
	\begin{abstract}
		
We show that a cyclic unitary process can extract work from the thermodynamic equilibrium state of an engineered quantum dissipative process.		
Systems in the equilibrium states of these processes serve as batteries, storing energy.
The dissipative 	process that brings the battery to the active equilibrium state is driven by an agent that couples the battery to thermal systems. The second law of thermodynamics imposes a work cost for the process; however, no work is needed to keep the battery in that charged state. We consider simple examples of these batteries and discuss particular cases in which the extracted work or the efficiency of the process is maximal. 
	\end{abstract}
	%\pacs{
	%05.70.Ln, % non-equilibrium and irreversible thermodynamics 
	%05.70.-a, % Thermodynamics
	%03.65.Yz %decoherence quantum mechanics
	%75.10.Pq %magnetic order spin chains
	%%05.60.Gg transport process quantum
	%}
	\maketitle
	
	%\section{introduction}
	\textit{Introduction:}
	In a dissipative process, the state of a system converges asymptotically to an invariant or steady state~\cite{Breuer}. 
	Thus, by engineering a dissipative-open-quantum-system dynamic, a system can be driven to a given target state.
	This idea has been explored in quantum computation, entanglement generation, and quantum sensing~\cite{a1,a2,a3,b1,b2,c1}, thanks to the control achieved on the coupling and coherent dynamics of multipartite quantum systems~\cite{JBarreiro,HWeimer,MMuller}.

	Another application relevant to future quantum technologies is charging a quantum battery. The process of charging a battery, i.e., storing energy in a quantum system for later use, has been studied in the context of unitary evolution while emphasizing the role of quantum correlations between its components~\cite{firstQB,otra bateria,quantacell1, quantacell2, campisibatt}. 
	Because the target state is given, charging a battery by a unitary process requires fine tuning between the initial state of the battery and the unitary process. This step is avoided by resetting the battery before the charging process.
	Alternatively, in quantum reservoir engineering, an agent can run a protocol without the need of any information of the battery state and charge it.
	The question that arises is how efficient such a process can be. 
	
	An energetically efficient engineered process should have no work cost to the agent if the battery has already achieved the charged state; otherwise, he dissipates the same currency that he wants to store.
	From a thermodynamic point of view, steady states of dissipative dynamics are either nonequilibrium states and dissipate energy or are {\it equilibrium} states, which are {\it dissipationless}. 
	Motivated by these considerations and recent progress in quantum thermodynamics~\cite{QTh1,Qth2,Qth3,MHP}, we consider the following question: 
	Can we engineer a dissipative process for a battery, involving auxiliary systems in the thermal Gibbs states all at the same temperature, such that its invariant state is an {\it equilibrium} state where energy can be stored and then extracted? We will show that, for systems with finite dimensional Hilbert space, this is indeed the case. We will characterize the charged equilibrium state by its ergotropy~\cite{ergotropy} and the charging process by its efficiency.
	
	The ergotropy of a state is the maximum work that can be extracted from it in a unitary cyclic process~\cite{ergotropy}. States with positive ergotropy are called \textit{active states} and those with vanishing, \textit{passive states}~\cite{passivity1,passivity2}. Equilibrium states reached by a system in a relaxation process are passive;
	otherwise, a perpetuum mobile of the second kind could be built in contradiction with the Kelvin Planck statement of the second law~\cite{termobook}. However, equilibrium states are not necessarily passive. It is possible to engineer processes with active equilibrium states. The second law implies that there is a work cost implementing the protocol that drives the system to the active equilibrium state~\cite{supmat}.

	When the optimal unitary cyclic process extracts the ergotropy, the battery is left in a corresponding passive state. We define the efficiency of the charging process by the ratio between the ergotropy of the equilibrium state and the work cost for the battery charging process from the corresponding passive state. If the battery is recycled after use, this will be the quantity of interest. 
	
	Usually, in thermodynamics, there is a tradeoff between the resource we are interested in and the efficiency of the process that produces it. 
	We will show 
	this tradeoff for the ergotropy of active equilibrium states. In particular, we find an engineered thermodynamic process that brings a battery to an equilibrium state with maximal ergotropy, as well as full population inversion but of low efficiency.
	A process of maximal efficiency with low ergotropy is illustrated with a two-qubit battery, and in~\cite{supmat}, we show that the efficiency generally goes to its maximal value only if the ergotropy goes to zero.

	\vspace{0.2cm}
	%\section{Thermodynamics of open quantum systems} 
	\textit{Thermodynamics of open quantum systems:} 
	We consider the following idealized scenario: we have a system of interest, the battery with a Hamiltonian $H_S$ and many copies of the same auxiliary system, each with the same Hamiltonian $H_A$.
	Initially, all systems are uncorrelated, and every copy of the auxiliary system is in the same temperature Gibbs thermal state $\omega_\beta(H_A)=e^{-\beta H_A}/\Tr[e^{-\beta H_A}]$, where $\Tr$ denotes the total trace. They play the role of a thermal bath. 
% Quality control editor: It is unclear what ``They'' refers to in the previous sentence. Please consider replacing it with ``The system copies'' or a more appropriate phrase.
The initial state of the battery is $\rho_S(0)$.
	Systems in the Gibbs state are easy to prepare by coupling them weakly to the environment with inverse temperature $\beta$ (we consider units such that $k_B=\hbar=1$). 
	To implement the desired dissipative dynamics,
	an agent couples the battery to one auxiliary system for a lapse of time $\tau$ and then to another system for a subsequent lapse of time $\tau$ and so on, turning on and off the interaction between the battery and different copies of the auxiliary systems, in a repeated interaction process~\cite{rep-int,Barra,Esposito rep. int.}. At the $n$th step, the battery interacts with the $n$th member of the auxiliary systems through the time independent potential $V$. This interaction vanishes at the initial $(n-1)\tau$, and final $n\tau$ times where the interaction is off and the total Hamiltonian is simply $H_S+H_A$. 
	The state of the battery at time $n\tau$ reads $\rho_S(n\tau)=\Tr_A[\rho_{\rm tot}(n\tau)]$, where $\Tr_X$ denotes a partial trace over the degrees of freedom of system $X$, in this case the $n$th auxiliary system,
	and
	\begin{equation} \label{ecc: evolucion total}
		\rho_{\rm tot}(n\tau) = U [\rho_S([n-1]\tau) \otimes \omega_\beta(H_A)] U^\dag,
	\end{equation}
	where $U=e^{i\tau(H_S+H_A+ V)}$ is the unitary time evolution operator for the composite system with the interaction on.
	Various dissipative processes are modeled in this way, such as streams of atoms moving across a quantum electrodynamic cavity~\cite{cavities} and a boundary driven system~\cite{rep-int2}, to mention a few.
	Recently, the importance of the time-dependent coupling for a proper thermodynamic description of these processes was discussed in~\cite{Barra} (see also~\cite{Chiara}), where the work performed by the agent due to switching on and off the interaction was taken into account.

	Let us briefly analyze the thermodynamics of the elementary process of duration $\tau$. The properties of the concatenated process are deduced from these, see ~\cite{StochPRE,StochPRE2} for details. 
	The energetics of the $n$th step is characterized by the switching work $W_n$, 
	and the heat $Q_n$ is the negative energy change of the $n$th auxiliary bath system \cite{Barra,Esposito rep. int.,Chiara}. These quantities read as follows: 
	\begin{eqnarray}
		&& W_n = \Tr[(H_S+H_A) (\rho_{\rm tot}(n\tau)-\rho_{\rm tot}([n-1]\tau))], \label{ecc: work}\\
		&& Q_n = -\Tr_A[H_A(\rho_A(n\tau)-\omega_\beta(H_A))], \label{ecc: heat}
	\end{eqnarray}
	where $\rho_A(n\tau)=\Tr_S[\rho_{\rm tot}(n\tau)]$.
	Their sum is the energy change of the battery (first law), calculated as
	\begin{equation}
		\Delta E_n = Q_n + W_n = \Tr[H_S(\rho_S(n\tau) - \rho_S([n-1]\tau))]. \label{ecc: energy change} \\
	\end{equation}
	Considering the von-Neumann entropy $S(\rho_S)=-\Tr_S[\rho_S\ln \rho_S]$, the entropy change of the battery in the $n$th step can be expressed as \cite{Esposito-NJP}
	\begin{equation}
		\Delta S_n =
		\Sigma_n + \beta Q_n\label{ecc: entropy change}\\
	\end{equation}
	where
	\begin{equation}
		%\Sigma_n = D(\rho_{\rm tot}(n\tau) || \rho_S(n\tau) \otimes \rho_A(n\tau)) + D(\rho_A(n\tau)||\omega_\beta(H_A)) \geq 0 \label{ecc: entropy production}
		\Sigma_n = D(\rho_{\rm tot}(n\tau) || \rho_S(n\tau) \otimes \omega_\beta(H_A)) \geq 0 \label{ecc: entropy production}
	\end{equation}
	is the entropy production with $D(a||b) \equiv \Tr[a\ln a]-\Tr[a\ln b]$ \textit{the relative entropy}, which is nonnegative for any density matrices $a$ and $b$~\cite{Breuer}. The inequality in Eq. (\ref{ecc: entropy production}) corresponds to the second law.
	
	In each time step, the battery evolves under the completely positive trace preserving map $\mathcal E(\rho)=\Tr_A[U\rho\otimes \omega_\beta(H_A) U^\dag]$, and we have characterized the thermodynamic properties for this elementary process. The engineered dissipative dynamics is obtained by concatenating this elementary process ($\mathcal E \circ \mathcal E\circ\cdots$).

	We engineer a $\mathcal E$ with a unique invariant state, $\pi=\mathcal E(\pi)$, attractive due to the contractive character of the relative entropy under the action of the map~\cite{Breuer}. Concatenating it a large number of times, every initial state of the battery will converge to $\pi$, i.e., $\lim_{n\to\infty} \mathcal E^n(\rho_S(0))=\pi$ $\forall \rho_S(0)$; the work, heat, and entropy produced in this process $\rho_S(0)\to\pi$ are the sum of the corresponding quantities, Eqs.~(\ref{ecc: work}), (\ref{ecc: heat}) and (\ref{ecc: entropy production}), for each step.

	We can distinguish two kinds of maps, maps with or without equilibrium~\cite{StochPRE,StochPRE2}. If the action of $\mathcal E$ over $\pi$ gives $\Sigma_n > 0$ then $\pi$ is a nonequilibrium steady state sustained by dissipated work ($W_n = -Q_n=\Sigma_n/\beta>0$) performed by the agent. Conversely, if the action of the map over $\pi$ gives $\Sigma_n=0$, then $\pi$ is an \textit{equilibrium state}. In equilibrium, the heat, work, and entropy production vanish ($W_n=Q_n=\Sigma_n=0$): no work is needed to sustain the state $\pi$. In this case, we say that $\rho_S(0)\to\pi$ is an {\it equilibration process}.
	The unitary time evolution operator $U$ of a map with equilibrium satisfies~\cite{StochPRE}
	\begin{equation} \label{ecc: map equilibrium condition}
		[U,H_0+H_A]=0
	\end{equation}
	with $H_0$ as an operator on the Hilbert space of the system, and the equilibrium state is $\pi=e^{-\beta H_0}/\Tr[e^{-\beta H_0}]=\omega_\beta(H_0)$. 
	
	Among the properties of maps with equilibrium, an important one is that the thermodynamic quantities Eqs. (\ref{ecc: work}), (\ref{ecc: heat}) and (\ref{ecc: entropy production}) can be written in terms of system operators only. Heat, work and entropy production take the form
	\begin{eqnarray}
		&&Q_n = \Tr_S[H_0(\rho_S(n\tau)-\rho_S([n-1]\tau)], \label{ecc: heat local} \\
		&&W_n= \Tr_S[(H_S-H_0)(\rho_S(n\tau)-\rho_S([n-1]\tau))], \label{ecc: work local}\\
		&&\Sigma_n = D(\rho_S([n-1]\tau)||\pi) - D(\rho_S(n\tau)||\pi) \geq 0. \label{ecc: entropy production local}
	\end{eqnarray}
	We see that if $H_0=H_S$, then $W_n=0$ and $\rho_S(0)\to \pi$ is {\it a relaxation process}. In this case, the equilibrium state is the passive Gibbs state. 
	If $H_0$ is an operator different from $H_S$, the equilibrium state $\pi=\omega_\beta(H_0)$ may be an active state and the \textit{equilibration process} $\rho_S(0)\to \pi$ has a total work cost $W=\sum_n W_n=\Tr_S[(H_S-H_0)(\pi-\rho_S(0))]$, as follows from Eq. (\ref{ecc: work local}). 
	Thus, a  condition necessary to have a charged battery in an equilibrium state is to engineer a map $\mathcal E$ with an equilibrium state $\omega_\beta(H_0)$ with $H_0\neq H_S$. 
	
	\vspace{0.2cm}
	%\section{ergotropy}
	\textit{Ergotropy:}
	To quantify the energy stored in a battery, we consider the ergotropy~\cite{ergotropy} 
	\begin{equation}
		{\mathcal W}(\rho_S)={\rm Max_u\, Tr}(H_S[\rho_S-u\rho_S u^\dag])
		\label{ineqW}
	\end{equation}
	of its state $\rho_S$. This is the maximal amount of work that can be extracted in a 
	unitary cyclic process, 
	where the state evolves unitarily with $u =\mathcal T_+ e^{-i\int dt(H_S+V_S(t))}$, and $V_S(t)$ is a time-dependent potential vanishing at the beginning and end of the process, 
	accounting for a cyclic external work source. 
	For {\it passive}~\cite{ergotropy,passivity2} states one has ${\mathcal W}(\rho_S)=0$. States are {\it active} if ${\mathcal W}(\rho_S)> 0$. 
	
	If we order the eigenvalues of $H_S=\sum_i E_i\ket{E_i}\bra{E_i}$ (assumed to be nondegenerate for simplicity) in increasing order, $E_1<E_2<\cdots <E_N$, and the eigenvalues of $\rho_S=\sum_i r_i\ket{r_i}\bra{r_i}$ in decreasing order, $r_1\geq r_2\geq\cdots\geq r_N$, then 
	the {\em ergotropy} of $\rho_S$ is given~\cite{ergotropy} by
	\begin{equation}
		\mathcal{W}(\rho_S)=\sum_{jk}r_jE_k(|\langle r_j|E_k\rangle|^2-\delta_{jk}).
		\label{ergotropy1}
	\end{equation}
	After the optimal work extraction process, the system is left in the corresponding passive state 
	\begin{equation}
		\label{passive}
		\sigma_{\rho_S}=
		\sum_j r_j\ket{E_j}\bra{E_j}.
	\end{equation}
	The ergotropy of $\rho_S$ can then be written as
	\begin{equation}
		\mathcal{W}(\rho_S)=\Tr_S[H_S(\rho_S-
		\sigma_{\rho_S})].
		\label{ergotropy2}
	\end{equation}
	
	\vspace{0.2cm}
	%\subsection{active equilibrium state}
	\textit{Condition for active equilibrium:}
	Let us obtain the conditions for an active {\it equilibrium state} $\omega_\beta(H_0)$.
	First, note that the equilibrium condition $[U,H_0+H_A]=0$ with $U=e^{-i\tau(H_S+H_A+ V)}$ is satisfied if $[H_0,H_S]=0$ and $[H_0+H_A, V]=0$.
	On the basis of common eigenvectors of the nondegenerate $H_S$ and $H_0$, the equilibrium state is
	\[
	\omega_\beta(H_0)=\sum_{i=1}^N\frac{e^{-\beta E^0_i}}{Z_0}\ket{E_i}\bra{E_i},
	\]
	and if a pair $(j,k)$ exists such that $(E_j-E_k)(E^0_j-E^0_k)\leq 0$, the state is active. Then, its ergotropy is extracted by a process described by a permutation unitary matrix $u$ associated to the permutation $p$ of $\{1,\cdots,N\}$ such that $E^0_{p_1}\leq \cdots \leq E^0_{p_N}$ leaving the battery in the passive state~\cite{ergotropy}
	\begin{equation}
		\sigma_{\omega_\beta(H_0)}=u\, \omega_\beta(H_0)u^\dag=
		\sum_{i=1}^N\frac{e^{-\beta E^0_{p_i}}}{Z_0}\ket{E_i}\bra{E_i}.
		\label{passive-3}
	\end{equation}
	
	Note that the total heat $Q_R$ and work $W_R$ obtained by Eqs. (\ref{ecc: heat local}) and (\ref{ecc: work local}) characterizing a recharging process $\sigma_{\omega_\beta(H_0)}
	\to\omega_\beta(H_0)$ are 
	\begin{eqnarray}
		\label{termoQtot}
		&&Q_R = \Tr_S[H_0 (\omega_\beta(H_0)-
		\sigma_{\omega_\beta(H_0)})], \\
		&&W_R = \Tr_S[(H_S - H_0)(\omega_\beta(H_0))-
		\sigma_{\omega_\beta(H_0)})],
		\label{termoWtot}
	\end{eqnarray} 
	and we see that the ergotropy of the state $\omega_\beta(H_0)$, obtained from Eqs. (\ref{ergotropy2}) and (\ref{passive-3}) is
	\[
	{\mathcal W}(\omega_\beta(H_0))=\sum_{i=1}^N (E_{p_i}-E_{i})\frac{e^{-\beta E_{p_i}^0}}{Z_0}
	\]
	and it is related to $W_R$ and $Q_R$ by
	\[
	W_R=\mathcal{W}(\omega_\beta(H_0))-Q_R.
	\]
	Note that $Q_R\leq 0$ (see Eq. (\ref{termoQtot})) because $\omega_\beta(H_0)$ is the state with minimum average $H_0$ among states with the same entropy~\cite{supmat}. It follows that $W_R\geq \mathcal{W}(\omega_\beta(H_0))\geq 0$, and thus, no perpetuum mobile of the second kind can be built. We quantify the efficiency of the charging process by the ratio 
	\[
	\eta\equiv\frac{\mathcal{W}(\omega_\beta(H_0))}{W_R}=1-\frac{|Q_R|}{W_R},\quad 0\leq \eta\leq 1.
	\]
	
	\vspace{0.2cm}
	%\subsection{High ergotropy}
	\textit{A protocol for active equilibrium:}
	A particularly interesting equilibrating processes with an active equilibrium state is obtained with 
	an interaction $V=\sum_\alpha S_\alpha\otimes A_\alpha$ where the system operators $S_\alpha$ and auxiliary bath operators $A_\alpha$
	satisfy $[H_S,S_\alpha]=\lambda_\alpha S_\alpha$ and $[H_A,A_\alpha]=\lambda_\alpha A_\alpha$. In this case, $[V,-H_S+H_A]=0$, i.e., we have $H_0=-H_S$, and 
	the corresponding process ${\mathcal E}$ has the equilibrium state 
	\[
	\omega_\beta(-H_S)=\sum_{i=1}^N \frac{e^{\beta E_i}}{Z_+}\ket{E_i}\bra{E_i}
	\]
	with $Z_+=\Tr e^{\beta H_S}$.

	Since different Hamiltonians with the same Bohr frequency spectrum $\{\lambda_\alpha\}$, are unlikely, the process should be engineered with auxiliary baths that are copies of the system, i.e., $H_S=H_A$.
	With this specific interaction $V,$ we have a process $\mathcal E$ with a remarkable thermodynamic equilibrium between a system in the state $\omega_\beta(-H_S)$ with copies of itself in the state $\omega_\beta(H_S)$. 
	Replacing $H_0=-H_S$ in Eqs.~(\ref{termoQtot}) and (\ref{termoWtot}), we see that in the recharging process, $Q_R=W_R/2={\mathcal W}$, see Eq.~(\ref{ergotropy2}), and the efficiency of this process is $\eta=1/2$.

	Since $E_j^0=-E_j$, the permutation $p$ that orders the spectrum $\{ E_j^0 \}$ in an increasing order is $i\to p_i=N+1-i$.
	Therefore, the ergotropy of $\omega_\beta(-H_S)$ is ${\mathcal W}=\sum_i (E_{N+1-i}-E_i)e^{\beta E_{N+1-i}}/Z_+$, which is positive, 
	and at low temperature, $\beta\to \infty$, it is the maximal value ${\mathcal W}=E_{N}-E_1$. 
	
	In general, ${\mathcal W}(\rho_S)$ is upper bounded~\cite{firstQB} by $\Tr[H_S(\rho_S-\omega_{\beta^*}(H_S))]$ with $\beta^*$ such that $S(\rho_S)=S(\omega_{\beta^*}(H_S))$; it is natural to ask 
	under what conditions ${\mathcal W}(\omega_\beta(-H_S))$ can saturate the bound. 
	We found that batteries with symmetric spectrum with respect to some energy value $\bar E$, i.e., $E_{N+1-i}=2\bar{E}-E_i$ saturate the bound. Indeed, in this case, 
	$E_{p_i}^0=E_{N+1-i}^0=-E_{N+1-i}=E_i-2\bar{E}$ and $Z_0=\sum_je^{-\beta E^0_j}=e^{2\beta\bar{E}}Z$ with $Z=\sum_j e^{-\beta E_j}$ as the canonical partition function. Thus, 
	the passive state of $\omega_\beta(-H_S)$ as given by Eq.~(\ref{passive-3}) is 
	\[
	\sigma_{\omega_\beta(-H_S)}=\sum_j \frac{e^{-\beta E^0_{N+1-j}}}{Z_0}\ket{E_j}\bra{E_j}
	%\]
	%\[
	% =\sum_j \frac{e^{\beta E_{N+1-j}}}{Z_0}\ket{E_j}\bra{E_j} 
	%=\sum_j \frac{e^{-\beta E_{j}}}{Z_0}\ket{E_j}\bra{E_j} 
	=\omega_\beta(H_S)
	\]
	i.e., the Gibbs state with the same temperature as the bath. 
	
\begin{figure}[htb]
		\includegraphics[width=7cm]{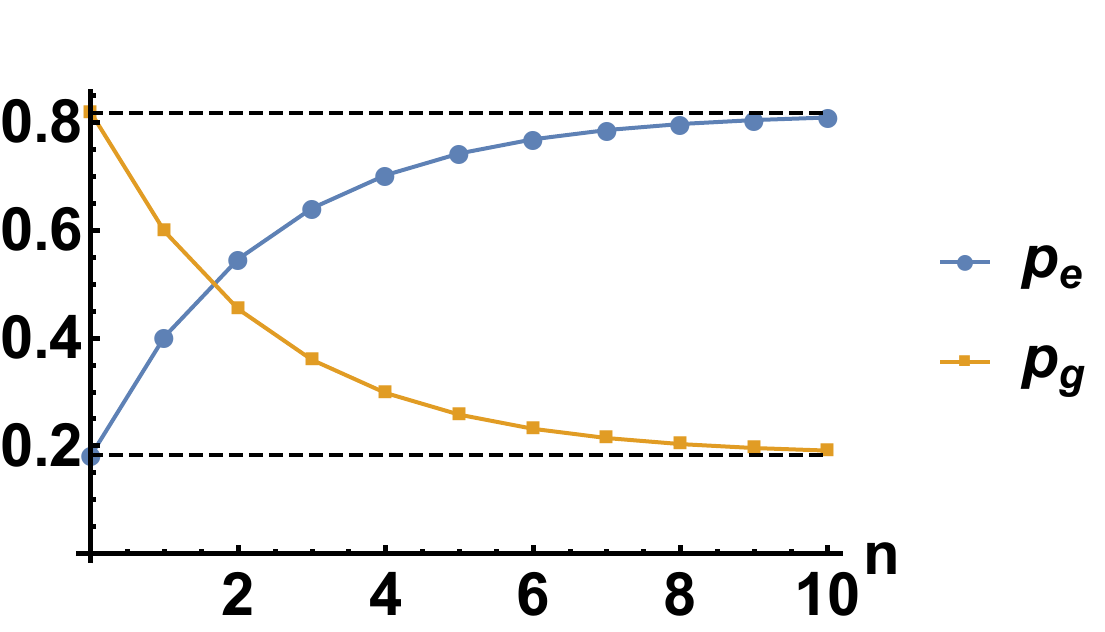}
		\includegraphics[width=7.5cm]{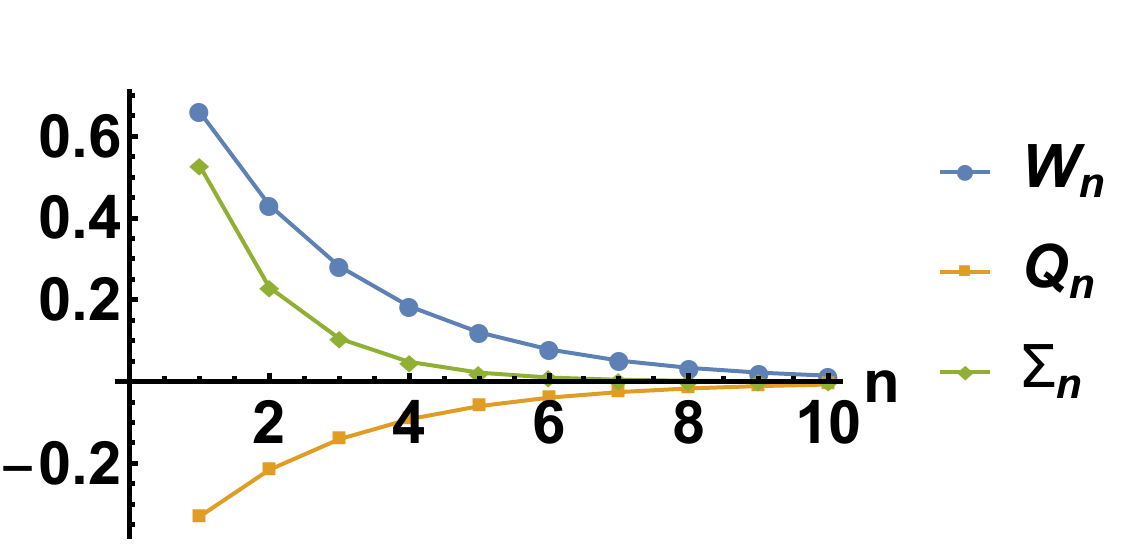}
		\caption{\label{figure1} \textbf{Population inversion in the equilibration process for the single-qubit battery.} \textbf{Upper panel}: Populations of the ground ($p_g$) and excited ($p_e$) states for each iteration step $n$. \textbf{Lower panel:} Work ($W_n$), heat flow ($Q_n$) and entropy production ($\Sigma_n$) for each step $n$. The initial state of the battery is a thermal state, that is, an identical state to that of the auxiliary baths. We consider for these plots $\tau=0.1$, $a=\sqrt{10}$, $h=1.5$ and $\beta=1$. The continuous lines are a guide for the eyes. }
	\end{figure}

	\vspace{0.2cm}
	%\section{Examples:}
	%\subsection{Single-qubit battery:}
	\textit{Single-qubit battery:}
	For our first example, we consider the battery and auxiliary systems all identical qubits, i.e., the battery Hamiltonian is $H_S=(h/2) \sigma_S^z$ (with a symmetric spectrum), and the auxiliary systems Hamiltonians are $H_A=(h/2) \sigma_A^z$, with $h>0$. The coupling between the system and the auxiliary qubit is 
	\[
	V=a(\sigma_S^+\sigma_A^++\sigma_S^-\sigma_A^-)
	\]
	and is such that $[\sigma_A^z-\sigma_S^z, V]=0$, i.e., $H_0=-H_S$. 
	The ergotropy of the battery in the equilibrium state $\omega_\beta(-H_S)$ is ${\mathcal W}=h\tanh \beta h/2$, which achieves the maximal value in the low temperature regime $\beta h\gg 1$. 
	In the upper panel of Fig.~\ref{figure1}, we plot the populations of the ground ($p_g$) and excited ($p_e$) states of the battery at each elementary step $n$ starting from the passive thermal state and in the lower panel, the thermodynamic quantities $W_n$, $Q_n$ and $\Sigma_n$. The population inversion is achieved, and the work cost goes to zero when the system reaches its equilibrium state.
	
	\vspace{0.2cm}
	%\subsection{two-qubit battery}
	\textit{Two-qubit battery:}
	With the previous protocol, we can achieve maximal ergotropy, especially in the low-temperature regime. We will now illustrate with another example that the maximal efficiency can also be achieved but with low ergotropy. In~\cite{supmat}, we show that this result is general. 
	
	We consider a two-qubit battery with Hamiltonian 
	\[
	H_S=\frac{h}{2}\left(\sigma^z_1+\sigma^z_2\right)+J\left(\sigma^x_1\sigma^x_2+\sigma^y_1\sigma^y_2\right).
	\]
	We take $2J>h>0.$ 
	If we consider 
	the process ${\mathcal E}$ obtained by coupling auxiliary systems of Hamiltonian $H_A=\frac{h}{2}\sigma_{A}^z$ to the battery of Hamiltonian $H_S$ with
	\[
	V=\sigma_{A}^x\sigma_1^x+\sigma_{A}^y\sigma_1^y,
	\]
	the equilibrium state is found to be $\omega_{\beta}(H_0)$ with $H_0=\frac{h}{2}\left(\sigma^z_1+\sigma^z_2\right)$, whose ergotropy is
	\[
	{\mathcal W}=(2J-h)\frac{\sinh \beta h}{1+\cosh \beta h}.
	\]
	The work done in the dissipative process $\sigma_{\omega_\beta(H_0)}\to \omega_\beta(H_0)$ that recharges the battery is
	\[
	W_R=2J \frac{\sinh \beta h}{1+\cosh \beta h}.
	\]
	We see that the efficiency $\eta = \mathcal W/W_R \to 1$ if $h\to 0$ for all $\beta$, yet, to have a finite ergotropy, one would need $\beta\sim{\mathcal O}(h^{-1})$. Note that if $\beta h \gg1$ but $h$ is small ${\mathcal W}\to 2J$, and for this system the state of maximum ergotropy is the pure state of maximal eigenenergy for which ${\mathcal W}=4J$. Details can be found in~\cite{supmat}.

	\vspace{0.2cm}
	%\section{Conclusions} \label{sec: conclusions}
	\textit{Conclusions:}
	We have shown that by engineering the coupling between a battery and auxiliary systems prepared in Gibbs thermal states, the battery undergoes a thermodynamic process that drives it into an {\it active equilibrium state}. The process has a work cost for the agent due to the coupling and decoupling between the battery and auxiliary systems. As a consequence, work can be extracted from the equilibrium state, but no perpetuum mobile of the second kind could be built. The notable aspect of our result is that because the charged state is an equilibrium state, the agent does not waste energy (work) once the battery is in the equilibrium state. One can thus consider that continuing the process once the battery is charged is a convenient way of protecting the charged battery. If a perturbation changes its state, the process will charge the battery again, spending energy only when this happens.

	We have characterized the activity of these equilibrium states by their ergotropy and the efficiency of the charging process; furthermore, we showed that in the low-temperature limit, either maximal ergotropy or efficiency could be obtained. 
	We observe a tradeoff between ergotropy and the efficiency $\eta$ of the process that produces it. Interestingly, we have found a dissipative process in which the equilibrium state of the system is $e^{\beta H}/Z_+$ while the environment is in the state $e^{-\beta H}/Z$.

	Finally, since all spin-spin 1/2 interactions are possible to implement with trapped ions~\cite{QGate}, the predictions for the qubit battery is testable with current experimental techniques. 
	%Optical pumping can be used to reset the state of the auxiliary qubit if many iterations are needed but as we show in [supplementary material] a few iterations produce an observable effect. 

	\section*{Acknowledgements}
	
	F.B. gratefully acknowledges comments from C. Lled\'o and the financial support of FONDECYT grant 1151390 and of the Millennium
	Nucleus ``Physics of active matter'' of the Millennium Scientific Initiative of the Ministry of Economy, Development and
	Tourism (Chile).

\end{document}

\def\mean#1{\left< #1 \right>}
%\preprint{}

\title{Dissipative charging of a quantum battery [supplementary material]}

\author{Felipe Barra\footnote{fbarra@dfi.uchile.cl}}
\affiliation{Departamento de F\'isica, Facultad de Ciencias F\'isicas y Matem\'aticas, Universidad de Chile, Santiago, Chile}

 \begin{abstract}
 \end{abstract}
%\pacs{
%05.70.Ln,  %	non-equilibrium and irreversible thermodynamics 
%05.70.-a,   %    Thermodynamics
%03.65.Yz %decoherence quantum mechanics
%75.10.Pq %magnetic order spin chains
%%05.60.Gg transport process quantum
%}
\maketitle

In this supplementary material we complement our discussion of passive states and 
provide a figure that illustrate the statement that the second law of thermodynamics does not forbid active states. We also give a few computations that we skipped in the main text. 
In the last section of this supplementary material, we generalize the repeated interaction protocol to take into account a lapse of time between the coupling of the battery to two consecutive auxiliary systems. We will see that this does not modify the main results.

\section{Passivity, ergotropy and the second law}

\subsection{Passivity and ergotropy}
The state $\rho$ is passive if for every unitary $u$
\begin{equation}
 %{\mathcal W}(\rho_S)={\rm Max_u\, Tr}(H_S[\rho_S-u\rho_S u^\dag])\leq 0
 {\rm Tr}(H[\rho-u\rho u^\dag])\leq 0,\quad\forall u
 \label{pasiveineq}
\end{equation}
where $H$ is the Hamiltonian of the system. The unitary $u$ represents a reversible process in which an agent couples a work source to the system. To account for the cyclic coupling with the work source, 
 we add to the Hamiltonian $H$ any time-dependent $V(t)$ for a time $\tau$, such that $V(t)$ is non-vanishing only when $0\leq t \leq \tau$. 
The corresponding evolution can be described by a unitary operator $u(\tau) ={\mathcal T}_+{\rm exp}(-i\int_0^\tau dt(H+V(t))$, where ${\mathcal T}_+{\rm exp}$ denotes the time-ordered exponential. By varying over all $V(t)$ we can generate any unitary operator $u = u\tau)$. In a unitary cyclic transformation the state of the system changes but not the Hamiltonian i.e. $(\rho,H)\to(\rho',H)$.
The inequality (\ref{pasiveineq}) expresses the fact that from a passive states work cannot be extracted in a cyclic operation. 

\vskip 0.3cm

The following is a well known 

\noindent {\bf Property} $\bigstar$: from all states $\rho$ with fixed entropy, $S(\rho)=-\Tr[\rho\ln\rho]$, the one with minimal $\Tr[H\rho]$ is the state $e^{-\beta H}/Z$ where $\beta$ is such that $S(\rho)=S(e^{-\beta H}/Z).$

\vskip 0.3cm

A direct consequence of {\bf property} $\bigstar$ is that the Gibbs state $\omega_{\beta}(H)=e^{-\beta H}/Z$ is passive, i.e., 
$\Tr[H(\omega_{\beta}(H)-\rho)]\leq 0$ where $\rho$ satisfies $S(\rho)=S(\omega_{\beta}(H))$.

The following three subsections discuss three important results.
The first analyzes the relation between passivity, the Kelvin-Planck formulation of the second law, and the states of thermodynamic equilibrium. The following two examine consequences of  {\bf property} $\bigstar$ that we mentioned in the main text.

For non-passive states, also called active states, the inequality (1) is not satisfied by all $u$ and the ergotropy of the active state is defined as ${\mathcal W}(\rho)={\rm max}_{u}{\rm Tr}(H[\rho-u\rho u^\dag])$. 
Note that for a given pair $(\rho,H)$ the ergotropy is fixed.

\subsection{The ergotropy can be optimized}

As discussed in the main text, 
once the ergotropy is extracted from $\rho$ the system is left in a corresponding passive state $\sigma_\rho$. 
As a consequence of the {\bf property} $\bigstar$, for a given pair $(\rho,H)$, the ergotropy is bounded $\mathcal{W}(\rho)=\Tr[H(\rho-\sigma_\rho)]\leq \Tr[H(\rho-\omega_{\beta^*}(H))]$, where 
$\beta^*$ is determined by the equality between the von Neumann entropies $S(\rho)=S(\sigma_\rho)=S(\omega_{\beta^*}(H)).$
The bound is achieved for some special pairs $(\rho,H)$ such that the corresponding passive state is $\sigma_\rho=\omega_{\beta^*}(H)$. As we discuss in the main text $(e^{\beta H}/Z_+,H)$ is such a pair if the spectrum of $H$ is symmetric, i.e., if there exists a value $\bar E$ such that if $E_i$ is eigenvalue then $2\bar{E}-E_i$ is also.  

\subsection{the second law does not forbid active equilirium states}
The passivity of the Gibbs state $\omega_{\beta}(H)=e^{-\beta H}/Z$ is a consequence of {\bf property} $\bigstar$ and 
 is often understood as a consequence of the second law of thermodynamics. The link between the passivity of the Gibbs state and the second law lies on the fact that a pure relaxation process involving no work-cost, i.e., a process 
such as connecting the system weakly to a thermal bath, will bring any state to the Gibbs state. 
In its Kelvin-Planck formulation:``There is no process whose sole effect is to extract heat from a body and transform it in work'' 
the second law, refers to a cycle in which both the Hamiltonian and the
state of the system (body) return to its original values.

Thus let us consider a thermodynamic cyclic process in which a system is initially and finally in thermodynamic equilibrium with a heat bath. Let us call this state $\rho$ and the Hamiltonian of the system $H$. 
The cyclic process is composed of two steps. In the first, a cyclic unitary process (as those involved in the definition of passivity) is performed on the system, i.e. $(\rho,H)\to (\rho',H)$.
The Hamiltonian returns to its original value, but the state has changed by the unitary transformation. 
In the second step, we couple the system again to the bath until thermal equilibrium is reached, i.e., $(\rho',H)\to (\rho,H)$, completing the cycle.

Consider that the second step of our cycle is a \textit{relaxation} process (no work cost) in which the system is weakly coupled to a heat bath, and thus $\rho$ is the Gibbs equilibrium state. Because the second law implies that the total work on the cycle cannot be negative, and the work in the second step vanish, the work performed on the first step, which equals the energy change $\Delta E$ of the system in this step, must be non negative, i.e., $\Delta E\geq 0$ and thus, Gibbs state is passive. 
This is the link between the passivity of the Gibbs state and the second law. See Fig. 1 (left panel).

Consider now that the second step of our cycle is an \textit{equilibration} process in which the system in contact with the bath reaches a state of thermodynamic equilibrium $\rho$ while some positive work $W>0$ is done during the process. If $\Delta E$ is the change in energy of the system in the first step, the second law restricts the total work performed on the cycle $\Delta E+W\geq 0$ and thus if $W>0$, work extraction $\Delta E<0$ form an equilibrium state is not forbidden by the second law.  See Fig. 1 (right panel). 
Equilibrations process with positive work cost $W>0$ are discussed in the main text.

\begin{figure}[htb]
\includegraphics[width=8.5cm]{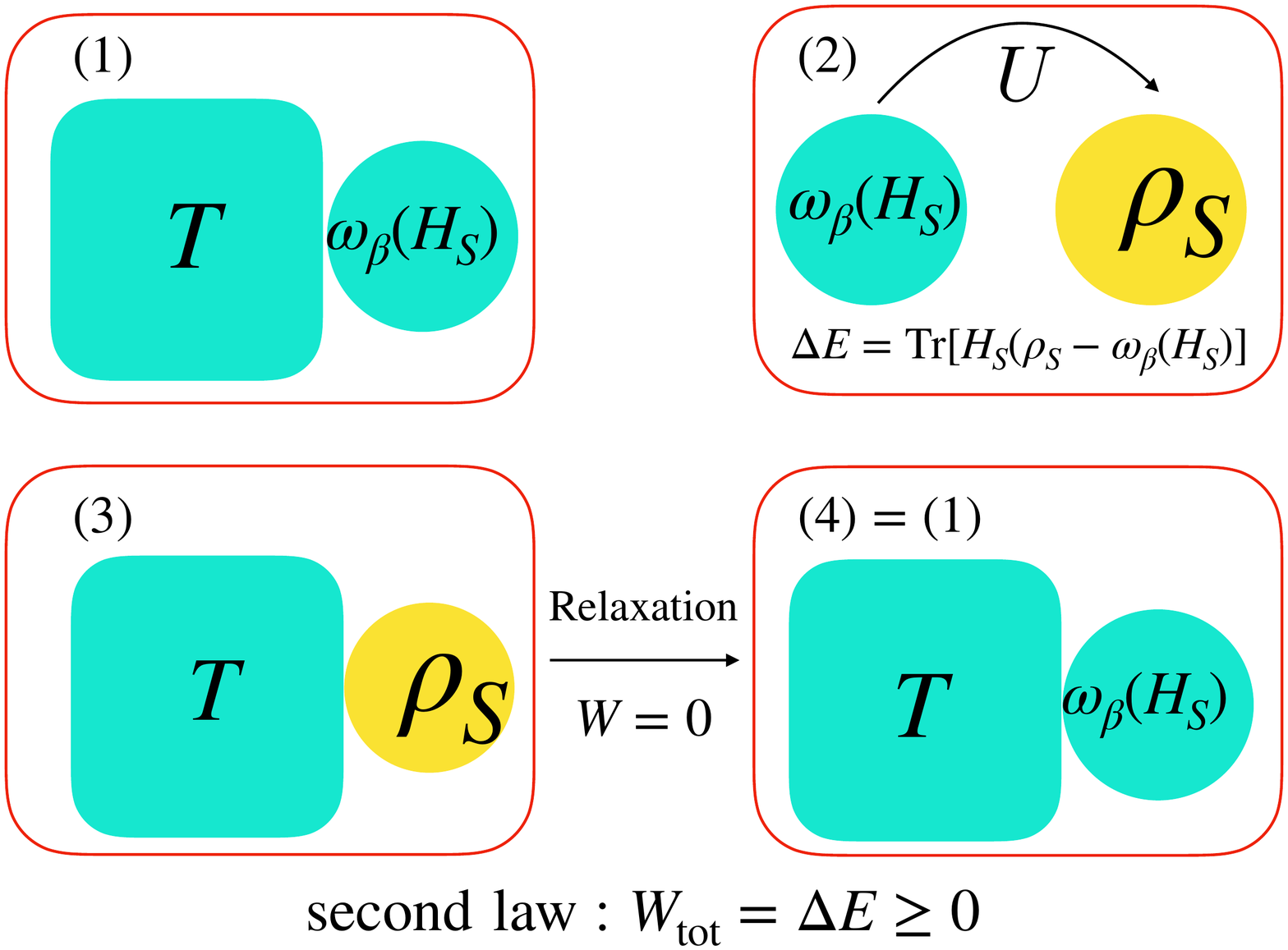}
\includegraphics[width=8.5cm]{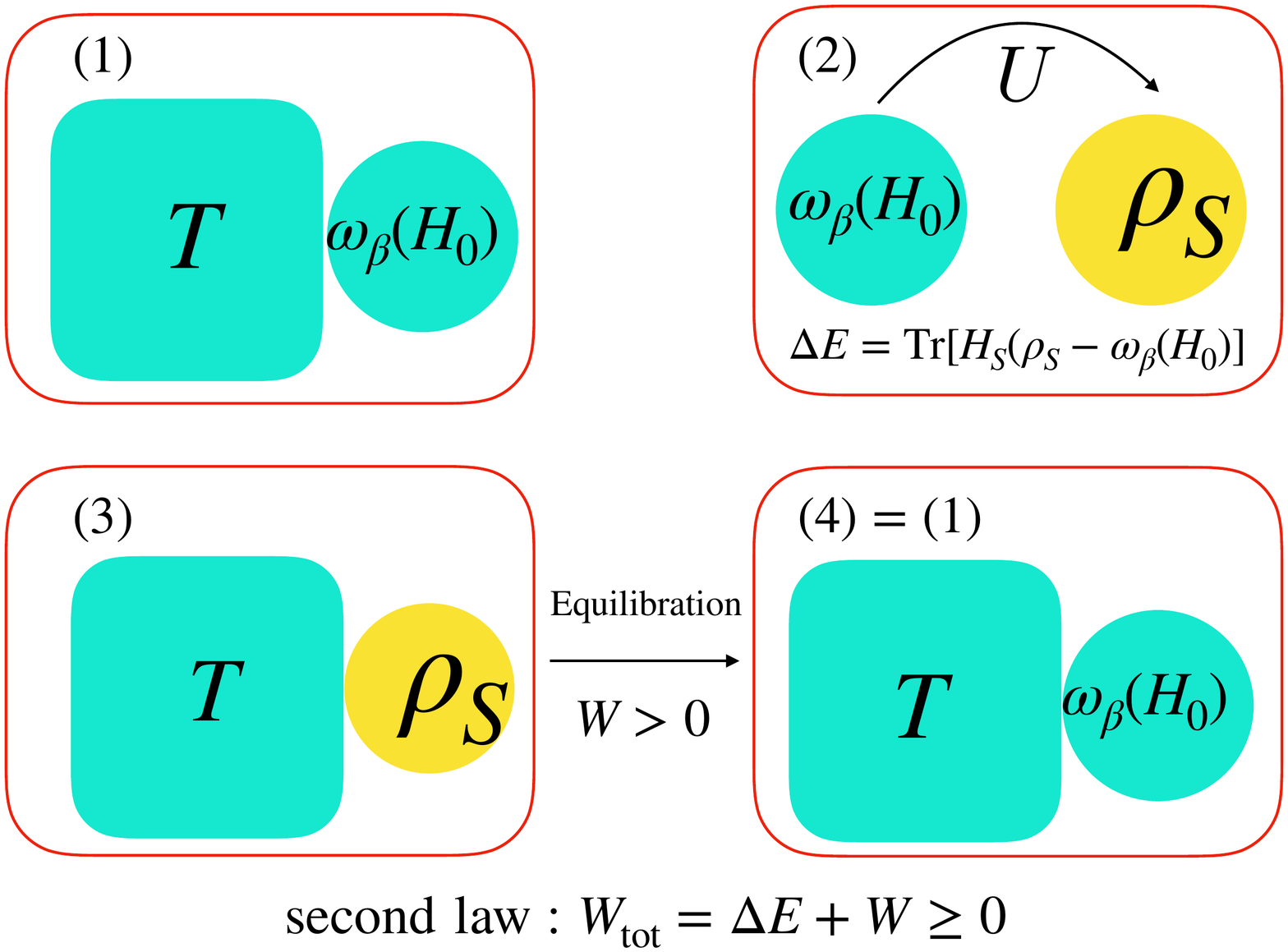}
\caption{{\bf Two thermodynamic cycles: Left panel: } (1) After relaxation, a system $S$ starts in the Gibbs state. (2) An agent performs work $\Delta E$ on the system. (3) The system is placed in contact with a heat bath. (3)$\to$(4) An undriven relaxation process brings the system back to the same Gibbs state of (1), closing the cycle. The second law implies that the total work performed by the agent in the cycle cannot be negative, i.e. $\Delta E\geq 0$.\newline
{\bf Right panel:} (1) After equilibration, a system $S$ starts in the $\omega_\beta(H_0)$ state. (2) An agent performs work $\Delta E$ on the system. (3) The system is placed in contact with a heat bath. (3)$\to$(4) The agent performs the equilibration process, performing work $W$ until the system reaches the state of (1), $\omega_\beta(H_0)$, closing the cycle. The second law implies that the total work performed by the agent in the cycle cannot be negative, i.e., $\Delta E+W\geq 0$, therefore the second law does not forbid $\Delta E=-{\mathcal W}(\omega_\beta(H_0))<0$.}
\end{figure}

\section{Recharging process and its efficiency}

The previous section discussed the general properties of passivity, ergotropy and equilibrium states. Here we consider the specific situation discussed in the main article. In particular, we derive some formulas and dwell on some arguments presented in the main article. We show that the efficiency of the charging process can go to 1, but the ergotropy vanish in the same limit. We also complement the proof that the charged equilibrium states we discuss cannot be used to build a perpetuum mobile of the second kind. Finally, we present some algebraic steps that are important for the charging protocol discussed in the main article.   

Because $[H_0,H_S]=0$, the equilibrium state 
\[
\omega_\beta(H_0)=\frac{e^{-\beta H_0}}{Z_0}=\sum_{j=1}^N \frac{e^{-\beta E^0_{j}}}{Z_0}\ket{E_{j}}\bra{E_{j}},
\]
can be ordered in the form $\rho=\sum_j r_j\ket{r_j}\bra{r_j}$ with $r_1\geq r_2\geq...\geq r_N$ by finding a permutation $p$ of $\{1,\cdots,N\}$ such that $E^0_{p_1}\leq E^0_{p_2}\leq...\leq E^0_{p_N}$, i.e, 
\[
\omega_\beta(H_0)=\sum_j \frac{e^{-\beta E^0_{p_j}}}{Z_0}\ket{E_{p_j}}\bra{E_{p_j}}.
\]
The passive state $\sigma_{\omega_\beta(H_0)}=\sum_j r_j \ket{E_{j}}\bra{E_{j}}$ is given by
\[
\sigma_{\omega_\beta(H_0)}=\sum_j \frac{e^{-\beta E^0_{p_j}}}{Z_0}\ket{E_{j}}\bra{E_{j}}.
\]
The unitary matrix that characterize the process of extracting the ergotropy $\sigma_{\omega_\beta(H_0)}=u\omega_\beta(H_0)u^\dag$ is given by the permutation matrix associated to $p$, i.e., its elements are $u_{ij}=\bra{E_i}u\ket{E_j}=\delta_{p_i,j}$. 
%Physically a permutation can be achieved by a Hamiltonian $H_S+V(t)$ such that the instantaneous energy levels $\{E_j(t)\}$ anticrosses twice as illustrated in {\color{blue}Figure S1}. Then by driving first slow (adiabatic crossing) and the fast, (diabatic crossing) the permutation is achieved. 
 
%\begin{figure}[htb]
%\includegraphics[width=8.5cm]{cruceLZ.pdf}
%\caption{Two avoided crossings, the system is driven fast for the first crossing and thus a diabatic transition exchanges the population of the levels. The system is slowly driven for the second avoided %crossing and thus an adiabatic transition preserves the populations of the levels. In this way a permutation of the population level is achieved.}
%\end{figure}
 
%
%permut DEM
%
%\[
%\omega=u^\dag \sigma u,\quad \sum\lambda_{p_j}\ket{\lambda_{p_j}}\bra{\lambda_{p_j}}= \sum\lambda_{p_j}u^\dag\ket{\lambda_{j}}\bra{\lambda_{j}}u
%\]
%entonces $\bra{\lambda_{i}}u=\bra{\lambda_{p_i}}$ por lo que $\bra{\lambda_{i}}u\ket{\lambda_j}=\bra{\lambda_{p_i}}\ket{\lambda_j}$ es decir $u_{ij}=\delta_{p_i,j}$

We have the following equalities for the ergotropy
\begin{equation}
{\mathcal W}(\omega_\beta(H_0))={\rm Tr}[H_S(\omega_\beta(H_0)-\sigma_{\omega_\beta(H_0)})]=\sum_{i=1}^N E_i\left(\frac{e^{-\beta E_i^0}}{Z_0}-\frac{e^{-\beta E_{p_i}^0}}{Z_0}\right)=\sum_{i=1}^N (E_{p_i}-E_{i})\frac{e^{-\beta E_{p_i}^0}}{Z_0}.
\label{Erg1}
\end{equation}

In the recharging process $\sigma_{\omega_\beta(H_0)}\to \omega_\beta(H_0)$ the work $W_R$ performed by the agent and the heat flow $Q_R$ from the bath are 
\begin{equation}
W_R={\rm Tr}[(H_S-H_0)(\omega_\beta(H_0)-\sigma_{\omega_\beta(H_0)})]=\sum_{i=1}^N(E_i-E_i^0)\left(\frac{e^{-\beta E_i^0}}{Z_0}-\frac{e^{-\beta E_{p_i}^0}}{Z_0}\right),
\label{Work1}
\end{equation}
\[
Q_R={\rm Tr}[H_0(\omega_\beta(H_0)-\sigma_{\omega_\beta(H_0)})]=\sum_{i=1}^NE_i^0\left(\frac{e^{-\beta E_i^0}}{Z_0}-\frac{e^{-\beta E_{p_i}^0}}{Z_0}\right),
\]
and they satisfy ${\mathcal W}(\omega_\beta(H_0))=W_R+Q_R$. The efficiency of the recharging process $\eta={\mathcal W}(\omega_\beta(H_0))/W_R$ satisfies $0\leq \eta\leq 1$ because ${\mathcal W}(\omega_\beta(H_0))\geq 0$,  $Q_R\leq 0$ and $W_R\geq 0$. 
These relations ruled out the existence of a perpetuum mobile of the second kind. In fact, we have the more general inequality 
\[
Q= \Tr_S[H_0 (\omega_\beta(H_0)-\rho_{S})]\leq 0
\]
for the heat in the process $\rho_S\to \omega_\beta(H_0)$ with $\rho_S$ the state of the system after work extraction. Indeed, work extraction is performed by a unitary process, $S(\rho_{S})=S(\omega_\beta(H_0))$ and {\bf property} $\bigstar$ implies, $\Tr_S[H_0 \omega_\beta(H_0)]\leq \Tr_S[H_0 \rho_S]$, therefore $Q\leq 0.$
In the main text we considered the case $\rho_{S}=\sigma_{\omega_\beta(H_0)}$ for which $Q=Q_R$ when the full ergotropy is extracted. The validity of the result for non-optimal unitary evolutions is important to rule out any perpetuum mobile. 

\section{most efficient battery and process}

In general, we see from Eqs. (\ref{Erg1}), middle expression, and (\ref{Work1}) that to have an efficient charging process $\eta\to 1$, the spectrum of $H_0$ must be very narrow (and thus dense). 
Indeed if the spectrum of $H_0$ is in a narrow energy band around some value $E^0$, i.e., 
$E_i^0=E^0+x_i\epsilon $ with $x_i\in[-1,1]$ and $\epsilon\ll E^0$. Then the heat 
\[
Q_R
%={\rm Tr}[H_0(\omega_\beta(H_0)-\sigma_{\omega_\beta(H_0)})]
=\sum_{i=1}^NE_i^0\left(\frac{e^{-\beta E_i^0}}{Z_0}-\frac{e^{-\beta E_{p_i}^0}}{Z_0}\right)=\epsilon\sum_{i=1}^Nx_i\left(\frac{e^{-\beta E_i^0}}{Z_0}-\frac{e^{-\beta E_{p_i}^0}}{Z_0}\right)\approx  \frac{e^{-\beta E^0}}{Z_0}\beta \epsilon^2 \sum_{i=1}^N x_i(x_{p_i}-x_i)
\]
is second order in $\epsilon$. 
The work $W_R$ and the ergotropy ${\mathcal W}(\omega_\beta(H_0))$ are both of first order in $\epsilon$ and equal to that order i.e. 
\begin{equation}
{\mathcal W}(\omega_\beta(H_0))=\sum_{i=1}^N E_i\left(\frac{e^{-\beta E_i^0}}{Z_0}-\frac{e^{-\beta E_{p_i}^0}}{Z_0}\right)\approx  \frac{e^{-\beta E^0}}{Z_0}\beta \epsilon \sum_{i=1}^N E_i(x_{p_i}-x_i)\approx W_R
\end{equation}
Therefore we see that $\eta\to 1$ as $\epsilon\to 0$ but ${\mathcal W}(\omega_\beta(H_0))\sim{\mathcal O}(\epsilon)$.

%Furthermore, we note that $e^{-\beta E_i^0}/Z_0\approx 1/N$ and that for a fully degenerate $H_0=E^0{\mathcal I}$ the equilibrium state is proportional to the identity and its ergotropy will vanish. 

\section{details for the two qubit example}

We consider a two qubit battery with Hamiltonian
%The 2 qubit Hamiltonian 
\[
H_S=\frac{h}{2}\left(\sigma^z_1+\sigma^z_2\right)+J\left(\sigma^x_1\sigma^x_2+\sigma^y_1\sigma^y_2\right)
\]
%have eigenvalues and eigenvectors that, in the basis defined by $\sigma^z\ket{0}=\ket{0}$ and $\sigma^z\ket{1}=-\ket{1}$, are  
%If we consider 
and the process ${\mathcal E}$ obtained by coupling auxiliary systems of Hamiltonian $H_A=\frac{h}{2}\sigma_{A}^z$ to the battery with
\[
V=\sigma_{A}^x\sigma_1^x+\sigma_{A}^y\sigma_1^y.
\]
One finds that the equilibrium state is $\omega_{\beta}(H_0)$ with $H_0=\frac{h}{2}\left(\sigma^z_1+\sigma^z_2\right)$, because $[H_0,H_S]=[V,H_0+H_A]=0$ (see ~\cite{StochPRE}). 

The eigenvalues and eigenvectors of $H_S$ and $H_0$ in the basis defined by $\sigma^z\ket{0}=\ket{0}$ and $\sigma^z\ket{1}=-\ket{1}$, are  
\begin{align}
E_3&=h,        & E^0_3&=h, &       \ket{E_3}&=\ket{00}\\
E_4&=2J,      & E^0_4&=0, &       \ket{E_4}&=(\ket{01}+\ket{10})/\sqrt{2}\\
E_1&=-2J,     & E^0_1&=0, &      \ket{E_1}&=(\ket{01}-\ket{10})/\sqrt{2} \\
E_2&=-h,       & E^0_2&=-h, &     \ket{E_2}&=\ket{11}
\end{align}

We take $2J>h>0$ such that $E_{i+1}>E_i$. The permutation that orders $E^0_{p_{i+1}}\geq E^0_{p_i}$ is $(p_1,p_2,p_3,p_4)=(2,1,4,3)$. 
Thus in the above basis, the equilibrium state is
\[
\omega_{\beta}(H_0)=\frac{e^{-\beta h} }{Z_0}\ket{E_3}\bra{E_3}+\frac{1}{Z_0}(\ket{E_1}\bra{E_1}+\ket{E_4}\bra{E_4})+\frac{e^{\beta h} }{Z_0}\ket{E_2}\bra{E_2}
\]
and the passive state for the system is 
\[
\sigma_{\omega_\beta(H_0)}=\frac{e^{\beta h} }{Z_0}\ket{E_1}\bra{E_1}+\frac{1}{Z_0}(\ket{E_2}\bra{E_2}+\ket{E_3}\bra{E_3})+\frac{e^{-\beta h} }{Z_0}\ket{E_4}\bra{E_4}
\]
where $Z_0=2+2\cosh(\beta h).$
The ergotropy of the equilibrium state ${\mathcal W}=\Tr[H_S(\omega_{\beta}(H_0)-\sigma_{\omega_\beta(H_0)})]$ is
\[
{\mathcal W}=(2J-h)\frac{\sinh \beta h}{1+\cosh \beta h}.
\]
The work done in the dissipative process $(\sigma_{\omega_\beta(H_0)},H_S)\to (\omega_\beta(H_0),H_S)$ that charge the passive state is
\[
W_r=2J \frac{\sinh \beta h}{1+\cosh \beta h}
\]
we see again that the efficiency $\eta\to 1$ if $h\to 0$ but ${\mathcal W}\to 0$ in that limit if $\beta$ is finite. If $\beta$ satisfies $\beta h \gg1$, the ergotropy ${\mathcal W}\to 2J$. Note that for this system the most ergotropic state is the pure state $\ket{E_4}$ for which ${\mathcal W}(\ket{E_4})=4J$.

\section{Generalized repeated interaction protocol}

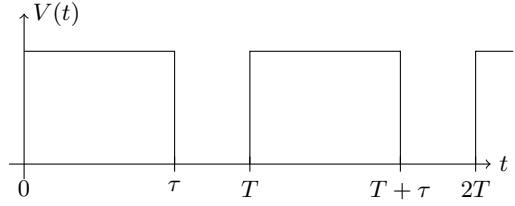
\begin{figure}
\begin{tikzpicture}
\draw[->](-0.2,0)--(6.2,0);
\draw[->](0,-0.2)--(0,2);
\node [right] at (6.2,0) {$t$};
\node [right] at (0,2) {$V(t)$};
\draw (0,1.5) -- (2,1.5) -- (2,0);
\node [below] at (0,-.1) {$0$};
\draw (2,-.1) -- (2, .1);
\node [below] at (2,-.1) {$\tau$};
\draw (3,-.1) -- (3, .1);
\node [below] at (3,-.1) {$T$};
\draw (3.,0)--(3.,1.5)--(5.,1.5)--(5.,0.);
\draw (5,-.1) -- (5, .1);
\node [below] at (5,-.1) {$T+\tau$};
\draw (6.,0)--(6.,1.5)--(6.5,1.5);
\draw (6,-.1) -- (6, .1);
\node [below] at (6,-.1) {$2T$};
\end{tikzpicture}
\caption{Time-dependent periodic potential. From $(n-1)T$ to $(n-1)T+\tau$, the interaction between the battery and the $n$th copy of the bath is on. From $(n-1)T+\tau$ to $nT$ if off.}
\label{figPot}
\end{figure}

In the dissipative process that we consider, many identical copies of the auxiliary bath system $A$, each with the same Hamiltonian $H_A$ and initially prepared in the same Gibbs state $\omega_\beta(H_A)$, interact sequentially with the battery for a time-lapse $\tau$ with the same coupling strength. Once the coupling is turned off, it will never interact again with the battery. In this section we will distinguish between one copy of the auxiliary system and another using the super-index $n$ for the state, $\rho_A^n(t)$, indicating that it is the $n$th copy of the auxiliary bath system. Similarly, we will write $H_A^n$ for the Hamiltonian of the $n$th copy of the bath, and $V^n$, the coupling operator between the system with the $n$th copy of the bath. 
When the agent controls a time-dependent interaction of the form
\[
V(t)=\sum_{n=1} V^n\Theta(t-[n-1]\tau)\Theta(n\tau-t),
\]
(where $\Theta(x)=1$ if $x\geq 0$, else $\Theta(x)=0$) a recursion for the system state is found:
\begin{equation} \label{ecc: system state recursion}
\rho_S(n\tau) = \Tr_n[U^n \rho_S((n-1)\tau) \otimes \omega_\beta^n(H_A^n) U^{n\dag}]={\mathcal E}( \rho_S((n-1)\tau)),
\end{equation}
with $U^n \equiv e^{-i\tau(H_S+H_B^n + V^n)}$ and $\Tr_n$ the trace over the system with Hamiltonian $H_A^n$~\cite{Attal, Barra}. 
The index $n$ in $U^n$, $H_B^n$, $V^n$ and $\rho_A^n$ distinguish the different Hilbert spaces $\mathcal H_A^n$ in  the total Hilbert space $\mathcal H = \mathcal H_S \otimes \mathcal H_A^1 \otimes \mathcal H_A^2 \otimes \cdots$, and it can be dropped because the operators have the same form. Therefore the map ${\mathcal E}$ in independent of $n$, i.e., 
$\Tr_n[U^n \rho \otimes \omega_\beta^n(H_A^n) U^{n\dag}]=\Tr_A[U\rho\otimes \omega_\beta(H_A)U^\dag]$ with the quantities in the right hand side defined in the main text. 
%We rewrite Eq.(\ref{ecc: system state recursion}) as 
%\[
%\rho_S(n\tau) ={\mathcal E}(\rho_S((n-1)\tau)) \equiv \Tr_A[U\rho_S((n-1)\tau)\otimes \omega_\beta(H_A)U^\dag].
%\]
Iterating $n$ times the recursion relation Eq.(\ref{ecc: system state recursion}) we have 
\begin{equation}
\rho_S(n\tau) ={\mathcal E}\circ\cdots\circ{\mathcal E}\rho_S(0).
\label{it1}
\end{equation}
Now, if a lapse of time $T-\tau$ occurs between switching off the interaction with the $n$th copy of the auxiliary bath and switching on the interaction with the $n+1$th copy, the time-dependent potential is (see Figure~\ref{figPot}) 
\begin{equation}\label{timedependentV}
V(t)=\sum_{n=1} V^n\Theta(t-[n-1]T)\Theta([n-1]T+\tau-t)
\end{equation}
and analogously one found
\begin{equation}
\label{it-free}
\rho_S(nT) =\left({\mathcal U}_{T-\tau}\circ{\mathcal E}\right)\circ\cdots\circ\left({\mathcal U}_{T-\tau}\circ{\mathcal E}\right)\rho_S(0)
\end{equation}
where ${\mathcal U}_t\rho=e^{-itH_S}\rho e^{itH_S}$.
To see this consider the initial condition $\rho_S(0)$ for the battery, so at $t=0$ the state of the full system, composed of the battery and the many copies of the auxiliary system, is 
\[
\rho_S(0)\bigotimes_{n=1} \omega^n_\beta(H_A^n).
\]
Because the battery and the first copy of the system are coupled for a lapse of time $\tau$ and then they are free for a lapse of time $T-\tau$, the state at time $T$ is 
\[
e^{-i(H_S+H_A^1)(T-\tau)}U^1\rho_S(0)\otimes \omega^1_\beta(H_A^1)U^{1\dag}e^{i(H_S+H_A^1)(T-\tau)}\bigotimes_{n=2} \omega^n_\beta(H_A^n)
\]
where we consider the invariance of the Gibbs state under the free evolution (in the lapse of time $T$) of the uncoupled auxiliary systems.  If we are interested in the state of the battery, we trace out all auxiliary systems obtaining
%\[
%\rho_S(T)=e^{-iH_S(T-\tau)}\Tr_1[e^{-iH_A^1(T-\tau)}U^1\rho_S(0)\otimes \omega^1_\beta(H_A^1)U^{1\dag}e^{iH_A^1(T-\tau)}]e^{iH_S(T-\tau)}
%\]
\[
\rho_S(T)=
%e^{-iH_S(T-\tau)}\Tr_1[U^1\rho_S(0)\otimes \omega^1_\beta(H_A^1)U^{1\dag}]e^{iH_S(T-\tau)}=
e^{-iH_S(T-\tau)}{\mathcal E}(\rho_S(0))e^{iH_S(T-\tau)}=\left({\mathcal U}_{T-\tau}\circ{\mathcal E}\right)\rho_S(0).
\]
Following references~\cite{Attal, Barra} one obtain the expression in Eq.(\ref{it-free}). Note that
if the initial condition is given by $\rho_S(t_i)$ with $t_i<0$ and the system evolves freely until the interaction with the first auxiliary system is turned on at $t=0$, then $\rho_S(0)= {\mathcal U}_{0-t_i}\rho_S(t_i)$. Similarly if the battery if free from $nT$ to $t_f\geq nT$ another map ${\mathcal U}_{t_f-nT}$ at the end of the chain in Eq.(\ref{it-free}) takes care of that.

An important remark is the following: Because we consider processes ${\mathcal E}$ with equilibrium states $\omega_\beta(H_0)=e^{-\beta H_0}/Z_0$ with $[H_S,H_0]=0$, the equilibrium state reached by the concatenated processes in Eq.(\ref{it1}) and Eq.(\ref{it-free}) is the same, therefore our results are also valid for generalized protocols with intermediate free evolutions. 

Consider that $m$ auxiliary systems have interacted with the battery between $t_i$ and $t_f$. The battery is free at these initial and final times. If $m$ is large, the battery may have reached a stationary state, but this is not important here. 
The energy change of the battery in this process
\begin{equation}
\Delta E=\Tr[H_S(\rho_S(t_f)-\rho_S(t_i))]=W+Q
\label{1stlaw}
\end{equation}
splits into two contributions: Work $W$ and heat $Q$. 
The work performed by the external agent corresponds to the energy change of the total isolated (battery plus auxiliary systems) system that evolves unitarily, 
\begin{equation}
\label{WW}
W=\int_{t_i}^{t_f}dt' \Tr[\rho_{\rm tot}(t')\dot H_{\rm tot}(t')]=\int_{t_i}^{t_f}dt' \Tr[\rho_{\rm tot}(t')\dot V(t')],
\end{equation}
where the upper dot indicates time derivative. The heat is minus the energy change of the auxiliary bath system
\begin{equation}
Q=-\sum_{n=1}^m \Tr[H_A^n(\rho_A^n([n-1]T+\tau)-\omega_\beta^n(H_A^n)],
\label{QQ}
\end{equation}
where $\rho_A^n([n-1]T+\tau)=\Tr_S[U^n\rho_S([n-1]T)\otimes\omega^n_A(H^n_A)U^{n\dag}].$

Let us show that adding them we get the battery energy change as expressed in Eq.(\ref{1stlaw}). 
Considering the time-dependent potential Eq.(\ref{timedependentV}) and using $\delta(x)=d\Theta(x)/dx$ in Eq.(\ref{WW}),we obtain
\[
W
%=\int_{t_i}^{t_f}dt' \sum_n \Tr[\rho_{\rm tot}(t')V^n](\delta(t-[n-1]T)-\delta(t-[n-1]T-\tau))
=\sum_{n=1}^m \Tr[\rho_{\rm tot}([n-1]T])V^n-\rho_{\rm tot}([n-1]T+\tau)V^n]
\]
from which we see that work is performed only switching on and off the interaction. Tracing out the systems that are not interacting at the corresponding time one gets
%\[
%W=\sum_{n=1}^m\left\{ \Tr_{S,n}[(\rho_S([n-1]T])\otimes\omega_\beta^n(H_A^n)V^n]-\Tr_{S,n}[U^n\rho_S([n-1]T])\otimes\omega_\beta^n(H_A^n)U^{n\dag}V^n]\right\}
%\]
%and 
after omitting the irrelevant index $n$, 
\begin{equation}
W=\sum_{n=1}^m\left\{\Tr[(\rho_S([n-1]T])\otimes\omega_\beta(H_A)V]-\Tr[U\rho_S([n-1]T])\otimes\omega_\beta(H_A)U^{\dag}V]\right\}.
%=-\sum_n\Tr[VU\rho_S([n-1]T)\otimes\omega_A(H_A)U^\dag]
\label{WW2}
\end{equation}
%The heat being minus the energy change of the auxiliary systems
%\[
%Q=-\sum_{n=1}^m \Tr[H_A^n(\rho_A^n([n-1]T+\tau)-\omega_\beta^n(H_A^n)]
%\] 

Forgetting the irrelevant index $n$ in Eq.(\ref{QQ}) for the heat we write
\begin{equation}
Q=-\sum_{n=1}^m \Tr[H_A(U\rho_S([n-1]T)\otimes\omega_A(H_A)U^\dag-\omega_\beta(H_A)]
\label{QQ2}
\end{equation}
and adding Eq.(\ref{QQ2}) and (\ref{WW2}) one obtains
%\[
%W+Q=\sum_n\left\{\Tr[(\rho_S([n-1]T])\otimes\omega_\beta(H_A)(V+H_A)]-\Tr[U\rho_S([n-1]T])\otimes\omega_\beta(H_A)U^{\dag}(V+H_A)]\right\}
%\]
%\[
%W+Q=\sum_n\left\{\Tr[(\rho_S([n-1]T])\otimes\omega_\beta(H_A)(V+H_A)]-\Tr[\rho_S([n-1]T])\otimes\omega_\beta(H_A)U^{\dag}(V+H_A)U]\right\}
%\]
\[
W+Q=\sum_n\left\{\Tr[(\rho_S([n-1]T])\otimes\omega_\beta(H_A)(V+H_A-U^{\dag}(V+H_A)U]\right\}.
\]
Noticing that $[V+H_A+H_S,U]=0$ implies $V+H_A-U^\dag (V+H_A)U=U^\dag H_SU-H_S$ we arrive a
%\[
%W+Q=\sum_n\left\{\Tr[(\rho_S([n-1]T])\otimes\omega_\beta(H_A)(U^{\dag}H_SU-H_S]\right\}=\sum_n\Tr[H_S({\mathcal E}(\rho_S([n-1]T))-\rho_S([n-1]T)]
%\]
\[
W+Q=\sum_n\Tr[H_S({\mathcal E}(\rho_S([n-1]T))-\rho_S([n-1]T)]=\sum_n\Tr[H_S(\rho_S([n-1]T+\tau)-\rho_S([n-1]T)].
\]
Considering that in the free evolution steps $\Tr[H_S(\rho_S([n-1]T+\tau)-\rho_S(nT)]=0$ we finally obtain the desired equality.
%because $\rho_S(nT)={\mathcal U}_{T-\tau}\rho_S([n-1]T+\tau)$ 
\[
W+Q=\sum_n\Tr[H_S(\rho_S(nT)-\rho_S([n-1]T)]=\Delta E
\]